\documentclass[final,3p,times,preprint]{elsarticle}

\usepackage{lineno,hyperref}
\usepackage{color}
\usepackage{siunitx}
\modulolinenumbers[5]

\journal{Journal of \LaTeX\ Templates}

\begin{document}

\begin{frontmatter}

\title{Development of Two-Dimensional Neutron Imager with a Sandwich Configuration}

\author[icepp]{Y. Kamiya\corref{corresp}}
\ead{kamiya@icepp.s.u-tokyo.ac.jp}
\cortext[corresp]{Corresponding author}

\author[KEK,SokenDai]{R. Nishimura}
\author[ShigaU]{S. Mitsui}
\author[LANL]{Z. Wang}
\author[LANL]{C. L. Morris}
\author[LANL]{M. Makela}
\author[LANL]{S. M. Clayton}
\author[LANL]{J. K. Baldwin}
\author[LANL]{T. M. Ito}
\author[RikkyoU]{S. Akamatsu}
\author[KEK,SokenDai]{H. Iwase}
\author[KEK]{Y. Arai}
\author[RikkyoU]{J. Murata}
\author[icepp]{S. Asai}

\address[icepp]{Department of Physics and International Center for Elementary Particle Physics, The University of Tokyo, 
Bunkyo-ku, Tokyo 113-0033, Japan}
\address[KEK]{High Energy Accelerator Research Organization (KEK), 
Tsukuba-shi, Ibaraki 305-0801, Japan}
\address[SokenDai]{The Graduate University for Advanced Studies (SOKENDAI), 
Hayama-ty$\hat{o}$, Kanagawa 240-0193 Japan}
\address[ShigaU]{Data Science and AI Innovation Research Promotion Center, Shiga University, 
Hikone-shi, Shiga 522-8522, Japan} 
\address[LANL]{Los Alamos National Laboratory, 
Los Alamos, New Mexico 87545, USA}
\address[RikkyoU]{Department of Physics, Rikkyo University, 
Toshima-ku, Tokyo 171-8501, Japan} 

\begin{abstract}
We have developed a two-dimensional neutron imager based on a semiconductor pixelated sensor,
especially designed for experiments measuring of a spatial and a temporal behavior of quantum bound states of ultra-cold neutrons.
Through these measurements, we expect to 
measure the ratio between the inertial and gravitational masses of neutrons
and to test the equivalence principle in the quantum regime.
As one of the principal neutron imagers,
we fabricated a sensor with a sandwich configuration, named $^{10}$B-INTPIX4-sw,
and tested its response to ultra-cold neutrons at the Los Alamos Neutron Science Center (LANSCE).
We observed simultaneous events on both sandwiching sensors without significant loss of detection efficiency.
The efficiency was evaluated to be about 16\%, relative to the $^{10}$B/ZnS reference detector.
The coincidence condition reduces its efficiency by a factor of about 3.
\end{abstract}

\end{frontmatter}

%=========================================
%=========================================
\section{Introduction}

Semiconductor pixelated sensors show great promise for tracking particle positions and trajectories.
These sensors can have pixel structures ranging from a few microns to tens of microns, 
allowing for high spatial resolution in particle and quantum beam imaging.
Their electrical controllability allows for the manipulation of electrical shutters, 
as well as the adjustment of image accumulation duration and timing of image readout.
This enables the incorporation of additional information such as time of flight into the image acquisition process.
By leveraging this additional degree of freedom, it becomes possible 
to determine third-dimensional positioning such as the depth of the particle emitter, 
and the particle's velocity/energy as its penetration power in the material it interacts with.
This extended functionality further advances the concept of an "imager."

We are conducting experiments to investigate the equivalence principle
within the framework of quantum theory, utilizing additional timing information\cite{PSI16, PSI19}. 
Experimental probes are gravitationally bound quantum states of Ultra-Cold Neutrons (UCNs)\cite{uGrav02,uGrav14}, 
a quantum system characterized by length and energy scales of 
\begin{eqnarray*}
z_0 \!&=&\! \left( \frac{1}{2 m_i m_g g} \right)^{1/3} \hspace{2mm} \sim 6~\unit{\micro\metre} \\
E_0 \!&=&\! \left( \frac{m_g^2 g^2}{2 m_i} \right)^{1/3} \hspace{2mm} \sim 0.6~\mathrm{peV}
\end{eqnarray*}
under the earth’s gravitational acceleration $g$. 
Here, natural units of $\hbar = 1$ are employed, 
where $m_i$ and $m_g$ represent the inertial and gravitational masses of neutrons, respectively. 
By analyzing the time evolution of the quantum system and simultaneously measuring these scales, 
one can evaluate the ratio $m_g/m_i$. Further details of the experiment will be published elsewhere.

Time-resolved neutron imager is an essential device for our experiment.
We have developed such an imager based on CMOS sensors using Silicon-on-Insulator (SOI) technology\cite{SOI}
and confirmed that the spatial resolution 
is achieved to be less than 4 \unit{\micro\metre} for fast neutrons\cite{b10int4}.

The measured resolution of 4 \unit{\micro\metre} represents the upper limit of the spatial resolution,
and it is considered that the imager's resolution is better than this.
This value primarily stems from imperfections in the beam alignment relative to the evaluation mask.
The inherent sources of uncertainty in determining neutron positions 
arise from electric noise and the finite travel distances of secondary particles emitted from a neutron-to-charged-particle converter.
To address or mitigate the uncertainty arising from particle ranges, 
we have developed a two-dimensional neutron imager with a sandwich configuration. 
In this paper, we present the first measurements of neutron events obtained using this new sensor setup.

%=========================================
%=========================================
\section{Principle of neutron imager}

\begin{figure}[thb]
\begin{center}
\includegraphics[width=0.98\linewidth]{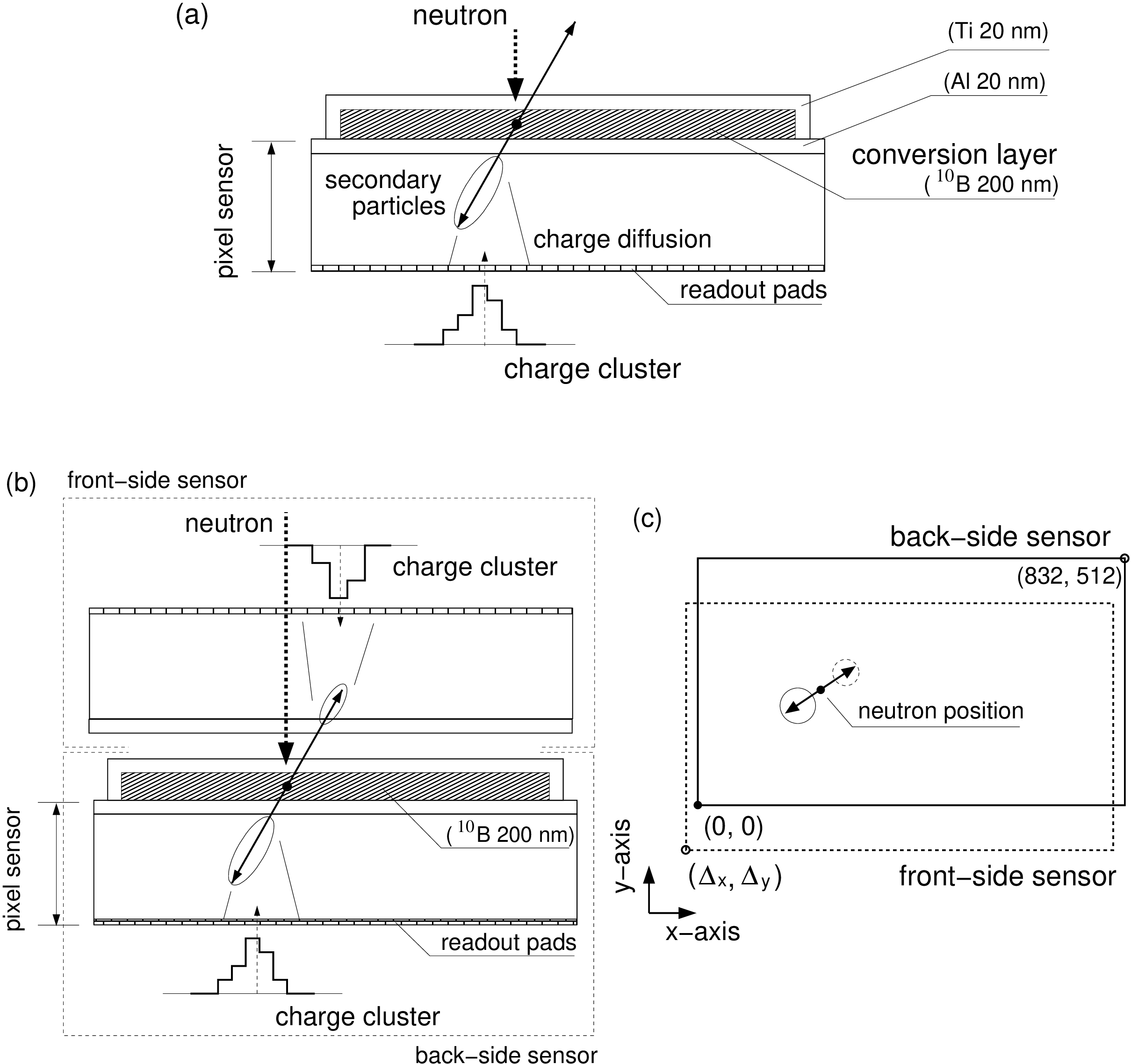}
\caption{(a) Schematic drawing of a previous neutron imager\cite{b10int4}. 
The process involves the conversion of a neutron into two charged particles through a nuclear reaction. 
One of these particles deposits its energy into the active volume of the pixelated sensor. 
The charges generated then diffuse and drift towards the readout pads, resulting in charge clusters due to neutron events. 
There is an intrinsic difference between the position 
where the neutron is injected and the charge-weighted mean of the cluster.
(b) An imager with a sandwich configuration.
The $^{10}$B conversion layer is formed on the back-side sensor.
(c) The front-side sensor is flipped and positioned with shifts, $\Delta x$ and $\Delta y$.
It is anticipated that the two corresponding charge clusters on both sides of the sensors 
will aid in a more accurate estimation of the neutron's position.}
\label{fig:imager}
\end{center}
\end{figure}

Our previous neutron imager\cite{b10int4} was fabricated as shown in Figure~\ref{fig:imager}a.
We employed an INTPIX4 SOI sensor\cite{int4Manual, int4Nishimura, int4Mitsui} as the base pixelated sensor.
Its specifications are listed in the Table ~\ref{table:int4}.
A thin ${}^{10}$B layer is deposited on the backplane of the sensor,
making the conversion of neutrons into charged particles through nuclear reactions as 
\begin{eqnarray*}
n + {}^{10}{\rm B} \!\!\!&\rightarrow&\!\!\! \alpha_{(1.47 ~{\rm MeV})} + {}^{7}{\rm Li}_{(0.84 ~{\rm MeV})} + \gamma_{(0.48 ~{\rm MeV})},\\
n + {}^{10}{\rm B} \!\!\!&\rightarrow&\!\!\! \alpha_{(1.78 ~{\rm MeV})} + {}^{7}{\rm Li}_{(1.01 ~{\rm MeV})}.
\end{eqnarray*}
The branching ratios are $94$ \% and $6$ \%, respectively.
Since $\alpha$ and $^7{\rm Li}$ are sufficiently heavier than the $\gamma$-ray energy,
they are emitted in opposite directions.
One of these particles enters the silicon active volume 
and generates electron-hole pairs along its trajectory.
These charges then diffuse, drift and are distributed among several readout nodes,
resulting in cluster images for neutron events.
%============================
%============================
\begin{table}
\centering
\begin{tabular}{lrr}
\hline
pixel size & 17 $\times$ 17 [$\unit{\micro\metre}^2$] \\
number of pixel & \hspace{3mm} 832 $\times$ 512 [pixel${}^2$] \\
active area & 14.1 $\times$ 8.7 [mm${}^2$] \\
readout time & 280 [ns/pixel] \\
wafer thickness & 300 [\unit{\micro\metre}] \\
\hline
\end{tabular}
\caption{Specification of INTPIX4 SOI sensor\cite{b10int4, int4Manual, int4Nishimura, int4Mitsui}.
Readout time is primarily determined by the settling time required 
to ensure that the effect of charge from the previous pixel is negligible.}
\label{table:int4}
\end{table}
%============================

Generally, cluster images from neutron events are isotropic, by comparing to other events such as from $\gamma$-ray radiation.
Therefore we can classify events by its particle types using a measure of the degree of isotropy of the cluster.

Two-dimensional neutron incident position, where the nuclear reaction occurs,
was estimated by a charge-weighted mean of pixel positions.
However, it is reasonable to assume that the charge-weighted mean 
estimates the position of the generated electron-hole cloud.
Therefore there is a fundamental difference between the true neutron position
and the cloud center, which is around the range of the secondary particle.
The ranges are estimated to be about 5 \unit{\micro\metre} for the $\alpha$ particle of 1.48 MeV 
and 2 \unit{\micro\metre} for the ${}^7$Li particle of 0.84 MeV in pure Si, respectively.

%=========================================
%=========================================
\section{Mitigation of the difference due to the range}

To mitigate the difference between the true neutron position and the estimated position,
we have develop an imager with a sandwich configuration,
as illustrated in Figure~\ref{fig:imager}b.
In this setup, the $^{10}{\rm B}$ neutron-to-charged-particle conversion layer is sandwiched between pixelated sensors. 
We anticipate achieving a more accurate estimation of the neutron position 
by individually estimating the position of each electron-hole cloud and averaging between them.

When identifying clusters originating from $\alpha$ or $^7{\rm Li}$ particles, 
we can estimate the position as the internal division point based on the representative range ratio.
This procedure is expected to yield a slightly more accurate estimation.
It's important to note that the ranges of $\alpha$ and $^7{\rm Li}$ particles in the active layer 
heavily depend on the amount of energy lost in the non-active volume they pass through beforehand. 
Well-controlled masses of non-active volumes are necessary for further quantitative discussions about this estimation scheme.

One of the drawbacks of this configuration is that neutrons must travel through the front sensor volume before being converted to secondary particles.
This situation could potentially reduce the detection efficiency, especially for very- or ultra-cold neutrons. 
To verify these effects, we conducted detector response tests using UCNs supplied at the Los Alamos Neutron Science Center (LANSCE).

%============================
%============================
\begin{table}
\centering
\begin{tabular}{r||r||r|r}
Imager & $^{10}$B/ZrS ref. detector \cite{b10ZrS} & $^{10}$B-INTPIX4-sw & $^{10}$B-INTPIX4-sw (single) \\
\hline\hline
measured flux [1/cm$^2$/s] & 1.8 & 0.16 & 0.42 \\
density monitor [arb. unit] & 691 & 379 & 379 \\
\hline
relative efficiency & 100 \% & 16 \% & 42 \% \\
\end{tabular}
\caption{Summary of the relative efficiency measurements.
The statistical uncertainty of $^{10}$B-INTPIX4-sw flux measurement was approximately 20\%.
The last column shows the relative efficiency 
when $^{10}$B-INTPIX4-sw is operated as a single-side sensor,
where the information from the front-side sensor were not utilized.
}
\label{table:eff}
\end{table}
%============================

\section{Signal from UCNs}

\begin{figure}[thb]
\begin{center}
\includegraphics[width=0.98\linewidth]{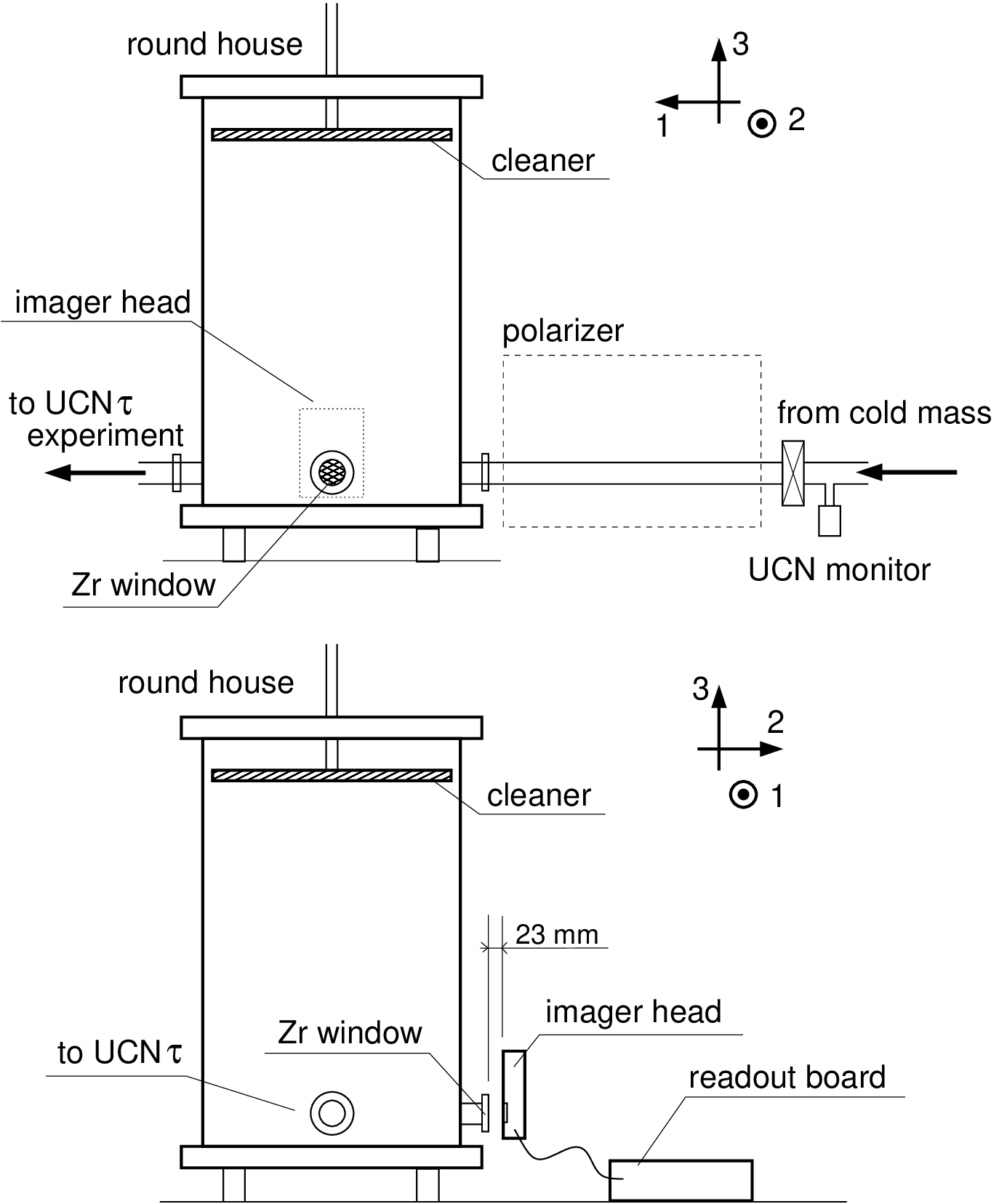}
\caption{Experimental setup for detection tests.
UCNs from the cold mass was filled in a buffer volume, referred to as the round house.
Cleaner was positioned select UCNs with energy around 100 neV.
The imager was positioned 23 mm away from the Zr window.
Direction 1 corresponds to the direction toward the UCN$\tau$ experiment\cite{ucnTau21, ucnTau22}, 
while direction 2 is to this imager test.
Direction 3 indicates the vertical axis.}
\label{fig:setup}
\end{center}
\end{figure}

Fig.~\ref{fig:setup} illustrates a schematic drawing of the experimental setup for the detection tests.
Neutrons generated in a spallation target of the 800 MeV LANSCE proton accelerator,
are cooled to the ultra-cold velocity of several meters per second using a solid deuterium cold mass.
These neutrons are then polarized by a superconducting magnetic polarizer 
and stored in a buffer volume called as the round house.
The density of ultra-cold neutrons is monitored at the gate valve situated in front of the round house.
Within the round house, there is a neutron cleaner designed to remove higher energy neutrons that could reach the ceiling position. 
Further details regarding the upstream equipments are provided in references\cite{ucnTau21, ucnTau22}.
UCNs pass through a Zr window, which is 25 \unit{\micro\metre} thick, into the air and then reach the testing imager.
The length of air between the window to the imager surface was 23 mm.
The setup is configured to select UCNs with an energy of around 100 neV.

The back-side sensor has the $^{10}$B neutron-charged-particle conversion layer.
The front-side sensor is flipped and fabricated 
with a shift of several tens of pixels in one direction relative to the back-side sensor.
The printed circuit board, on which the sensors are mounted, has an elongated hole to accommodate the passage of bonding wires.
In Figure~\ref{fig:imager}c, the axis for pixel coordinates is defined,
where $\Delta_x$ and $\Delta_y$ represent the relative shifts.
The origins of each sensor are at the bottom left corner.

We follow similar procedures for pedestal estimation/correction, charge clustering, 
neutron event selection, and the position estimation as in our previous paper\cite{b10int4}.
Pedestal estimation for each image is carried out by averaging the successive 10 images before and after. 
After the pedestal correction, we identify a local maximum pixel to serve as the cluster seed, 
which exceeds the threshold level of 300 Analog-to-Digital-Converter (ADC) channels. 
Since energy calibration for the imager has not been performed yet, 
the pixel charge is counted in units of ADC channels. 
A 7 pixels $\times$ 7 pixels region around the local maximum pixel is defined as the event cluster. 
We set a second threshold level at 120 ADC channels and count the number of pixels that exceed this threshold. 
If only one pixel exceeds the second threshold within the cluster area, 
the event is considered to be from sources other than a neutron event, 
such as X-ray or electric noise, and is therefore rejected as a single-pixel cluster.

The features of the cluster shape, cluster charge ($m_0$), charge-weighted mean ($\vec{m}_1$) 
and the second order center moment ($\vec{m}_2$) are defined as
\begin{eqnarray*}
m_0 &=& \sum_{i=1}^{7\times7} q_i, \\
\vec{m}_1 &=& \frac{1}{m_0} \sum_{i=1}^{7\times7} q_i \vec{r}_i\\
\vec{m}_2 &=& \frac{1}{m_0} \sum_{i=1}^{7\times7} q_i (\vec{r}_i - \vec{m}_1)^2,
\end{eqnarray*}
where $q_i$ represents the charge of the $i$th pixel, and $\vec{r}_i = (x_i, y_i)$ denotes the position of the pixel.

\begin{figure}[thb]
\begin{center}
\includegraphics[width=0.98\linewidth]{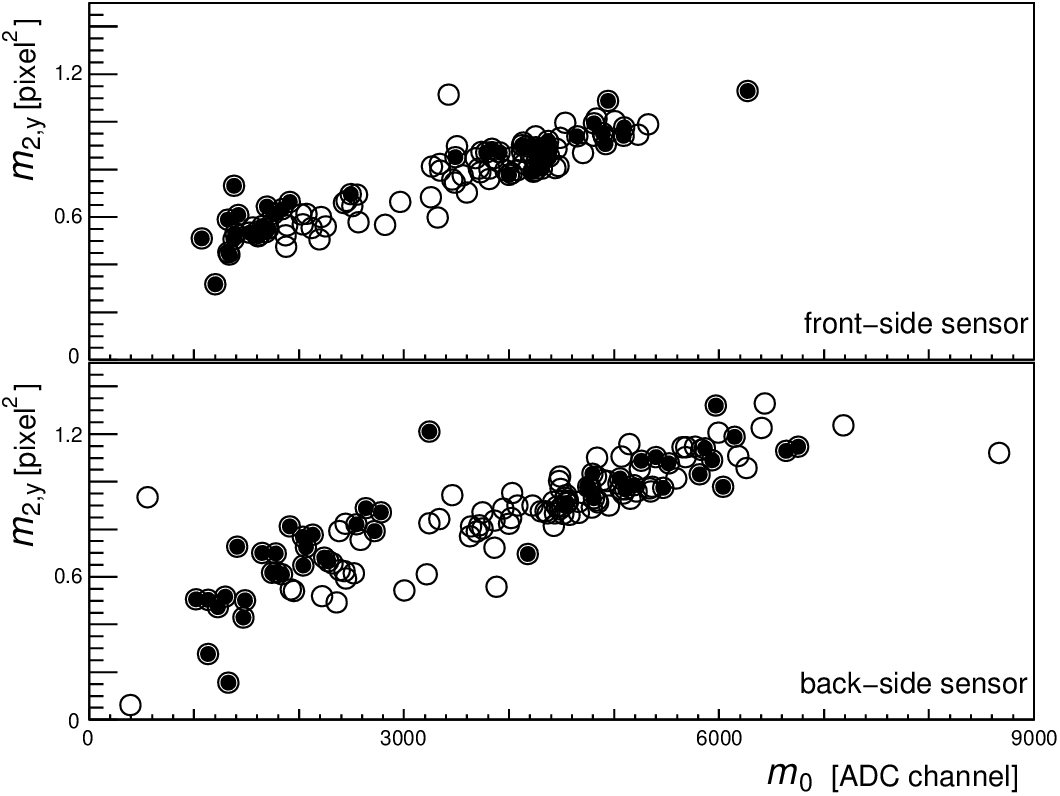}
\caption{The scatter plots of $m_{2,y}$ vs. $m_0$ for
the front-side sensor (above) and the back-side sensor (below).
Events meeting the coincidence condition on both side sensors are represented by filled circles.
Events with higher or lower than $m_0 \!\sim\! 3000$ ADC channels 
corresponds $\alpha$- or ${}^7$Li-origin events, respectively.
The acceptance for the paired-up events was about 1/3.
}
\label{fig:scatt}
\end{center}
\end{figure}

Neutron event shows a characteristic distribution on $m_{2,y}$ ($m_{2,x}$) vs $m_0$ scatter plot\cite{b10int4},
where $m_{2,y}$ and $m_{2,x}$ are $y$ and $x$ component of the vector $\vec{m}_2$, respectively.
Fig.~\ref{fig:scatt} illustrates the measured events as circles for the front-side (above) and the back-side sensors (below).
Two categories, those above and below $m_0 \sim 3000$,
represent $\alpha$- and $^{7}$Li-origin clusters, respectively.
Among these scatter plots, events measured simultaneously by the sensors are depicted as filled circles.
The number of events reduces from 111 events to 46 events (from 133 events to 46 events) 
for the front-side (for the back-side) sensor, respectively, by applying the coincidence condition.

\begin{figure}[thb]
\begin{center}
\includegraphics[width=0.98\linewidth]{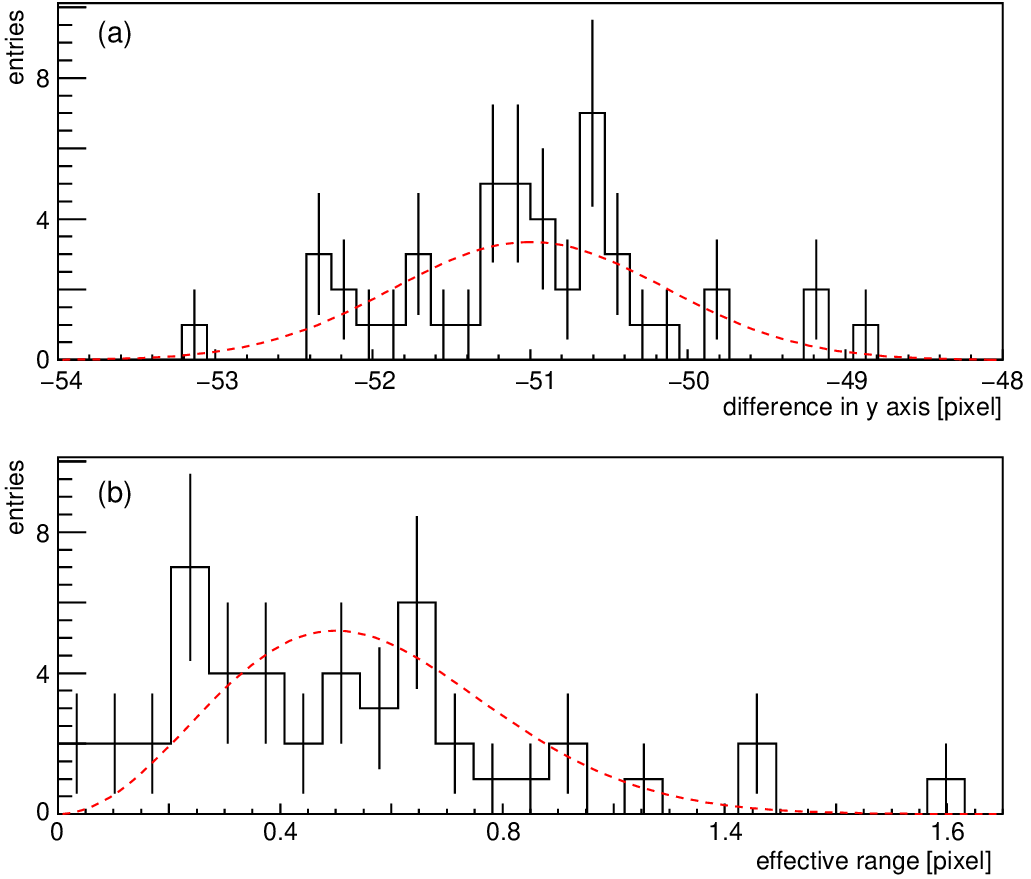}
\caption{(a) Position difference of paired clusters in $y$-axis.
Error bars correspond to statistical uncertainty, represented by the square root of each entry in the bin.
The dashed line indicates a fitted Gaussian function, where the mean value of -50.99 represents the amount of shift 
of the front-side sensor with respect to the back-side sensor.
(b) The distribution shows the half of position difference of paired clusters in the $y$-$x$ 2-dimensional plane.
The dashed line represents a Maxwellian fit. The most probable averaged range was 0.5 pixels.
}
\label{fig:diff}
\end{center}
\end{figure}

\begin{figure}[thb]
\begin{center}
\includegraphics[width=0.98\linewidth]{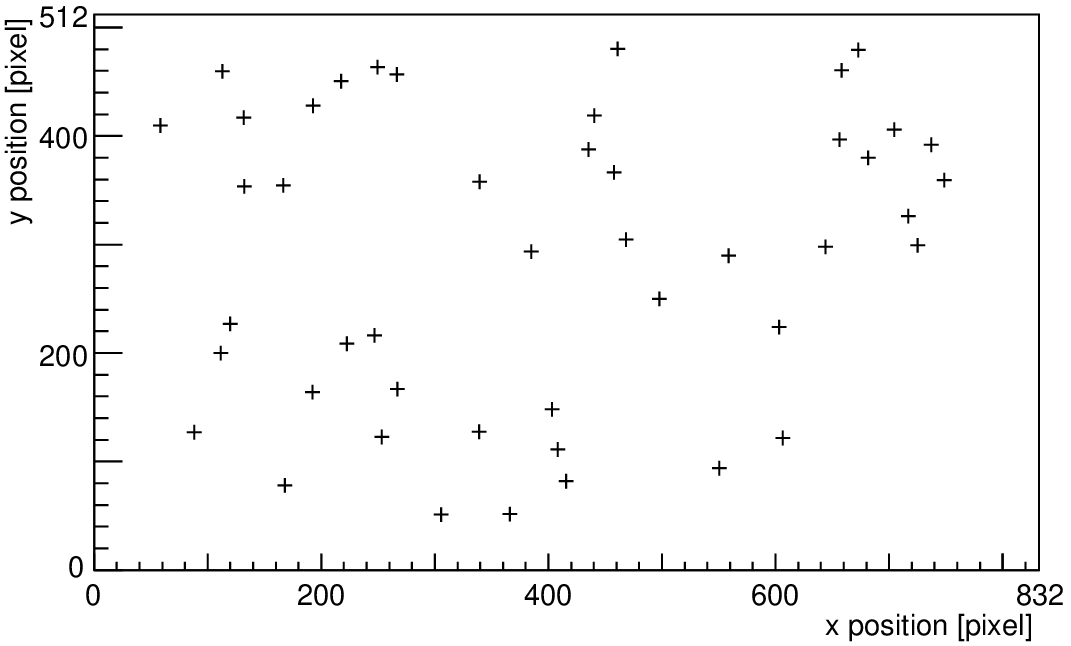}
\caption{Measured position of UCN events with the $^{10}$B-INTPIX4-sw imager.
The area with $100 <$ $x$-position $< 700$ and $100 <$ $y$-position $< 450$ 
was used to estimate the measured flux.}
\label{fig:position}
\end{center}
\end{figure}

One can estimate the relative shift of two sensors, $\Delta_x$ and $\Delta_y$,
by analyzing the differences in event coordinates.
Fig.~\ref{fig:diff}a illustrates the coordinate differences of the paired clusters along the $y$ axis, 
where $\Delta_y$ was estimated to be $\Delta_y = -50.99$ pixels based on the mean values of the fitted Gaussian distribution.
Similarly, $\Delta_x$ wes estimated to be $\Delta_x = -0.04$ pixels.
The widths of these Gaussians reflect the ranges of the secondary particles and uncertainties in positioning.
By averaging the two particle, effective projected range is shown as Figure~\ref{fig:diff}b,
estimated to be about 0.50 pixels as the most probable range, 
assuming the distribution follows a Maxwellian distribution.
The charge sum and charge difference of the paired clusters reflect
the $Q$ value of the nuclear reaction and the mass difference of the secondary particles, respectively.
The estimated neutron positions are shown in Figure. \ref{fig:position}.
For area with $x$-position greater than around 750 pixels, the converter layer is not coated, 
resulting in no sensitivity to neutrons.

\section{Detection efficiency}

The UCN flux at the position of the testing imager was measured 
by the $^{10}$B/ZrS scintillation detector\cite{b10ZrS} to be 1.8 1/cm$^2$/s.
The corresponding UCN averaged density monitored at the gate valve position was 691 in arbitrary unit.
We estimate the relative detection efficiencies based on this detector below.

The results are summarized in Table~\ref{table:eff}.
Measured flux by the ${}^{10}$B-INTPIX4-sw was 0.16 1/cm${}^{2}$/s,
with an averaged density of 379.
These results indicate that the relative efficiency of the coincidence measurement is approximately 16\%.

The back-side sensor measured a neutron flux of 0.46 1/cm${}^{2}$/s,
corresponding to a relative efficiency of 42\%.
The efficiency is reduced to 1/3 due to the coincidence condition of the sandwiching sensors, 
attributed to the acceptances of the secondary particles.
Quantitative discussion of acceptance requires control over sensor placement 
and material budget of non-active volume, which is beyond the scope of this paper.

\section{Summary}
A new two-dimensional neutron imager, $^{10}$B-INTPIX4-sw, was developed in a sandwich configuration,
with a measured relative efficiency of approximately 16\% for UCNs.
Operating the imager with only the back-side sensor, without the coincidence condition, yielded a relative efficiency of about 42\%.
The acceptance for the coincidence condition was approximately 1/3.

\section*{Acknowledgements}
We appreciate Toshinobu Miyoshi, formerly affiliated with KEK \!-\! Electronics System Group,
for his significant contributions at the early stages of the development.
We wish to thank Yoshio Mita, Ayako Mizushima, and the other members of the Takeda Sentanchi super clean room
at the University of Tokyo for operating the laboratory. 
This work received partial support by JSPS KAKENHI Grant Number 23H00106, 18H04343, 18H01226, and 17H05397.
The fabrication and testing of the sandwich structure prototypes were supported 
by the TIA Kakehashi framework in 2023, 2022, and 2021.

\bibliography{HSTD13}

\providecommand{\noopsort}[1]{}\providecommand{\singleletter}[1]{#1}%
\begin{thebibliography}{10}
\expandafter\ifx\csname url\endcsname\relax
  \def\url#1{\texttt{#1}}\fi
\expandafter\ifx\csname urlprefix\endcsname\relax\def\urlprefix{URL }\fi
\expandafter\ifx\csname href\endcsname\relax
  \def\href#1#2{#2} \def\path#1{#1}\fi

\bibitem{PSI16}
Y.~Kamiya, {Search for New Gravity-like Interactions and Test of the
  Equivalence Principle using Slow Neutrons}, The Physics of Fundamental
  Symmetries and Interactions - PSI2016, Switzerland
  https://indico.psi.ch/event/3914/ (2016).

\bibitem{PSI19}
Y.~Kamiya, {Testing the Weak Equivalence Principle using Grav- itationally
  Bound Quantum States of Ultracold Neutrons}, The Physics of Fundamental
  Symmetries and Interactions - PSI2019, Switzerland
  https://indico.psi.ch/event/6857/ (2019).

\bibitem{uGrav02}
V.~V. Nesvizhevsky, H.~G. Börner, A.~K. Petukhov, H.~Abele, S.~Baeßler, F.~J.
  Rueß, T.~Stöferle, A.~Westphal, A.~M. Gagarski, G.~A. Petrov, A.~V.
  Strelkov, {Quantum states of neutrons in the Earth's gravitational field},
  Nature 415~(6869) (2002) 297--299.
\newblock \href {https://doi.org/10.1038/415297a} {\path{doi:10.1038/415297a}}.

\bibitem{uGrav14}
G.~Ichikawa, S.~Komamiya, Y.~Kamiya, Y.~Minami, M.~Tani, P.~Geltenbort,
  K.~Yamamura, M.~Nagano, T.~Sanuki, S.~Kawasaki, M.~Hino, M.~Kitaguchi,
  {Observation of the Spatial Distribution of Gravitationally Bound Quantum
  States of Ultracold Neutrons and Its Derivation Using the Wigner Function},
  Physical Review Letters 112~(7) (2014) 071101.
\newblock \href {https://doi.org/10.1103/physrevlett.112.071101}
  {\path{doi:10.1103/physrevlett.112.071101}}.

\bibitem{SOI}
Y.~Arai, T.~Miyoshi, Y.~Unno, T.~Tsuboyama, S.~Terada, Y.~Ikegami, R.~Ichimiya,
  T.~Kohriki, K.~Tauchi, Y.~Ikemoto, Y.~Fujita, T.~Uchida, K.~Hara, H.~Miyake,
  M.~Kochiyama, T.~Sega, K.~Hanagaki, M.~Hirose, J.~Uchida, Y.~Onuki, Y.~Horii,
  H.~Yamamoto, T.~Tsuru, H.~Matsumoto, S.~Ryu, R.~Takashima, A.~Takeda,
  H.~Ikeda, D.~Kobayashi, T.~Wada, H.~Nagata, T.~Hatsui, T.~Kudo, A.~Taketani,
  T.~Kameshima, T.~Hirono, M.~Yabashi, Y.~Furukawa, M.~Battaglia, P.~Denes,
  C.~Vu, D.~Contarato, P.~Giubilato, T.~Kim, M.~Ohno, K.~Fukuda, I.~Kurachi,
  M.~Okihara, N.~Kuriyama, M.~Motoyoshi, {Development of SOI pixel process
  technology}, Nuclear Instruments and Methods in Physics Research Section A:
  Accelerators, Spectrometers, Detectors and Associated Equipment 636~(1)
  (2011) S31--S36.
\newblock \href {https://doi.org/10.1016/j.nima.2010.04.081}
  {\path{doi:10.1016/j.nima.2010.04.081}}.

\bibitem{b10int4}
Y.~Kamiya, T.~Miyoshi, H.~Iwase, T.~Inada, A.~Mizushima, Y.~Mita, K.~Shimazoe,
  H.~Tanaka, I.~Kurachi, Y.~Arai, {Development of a neutron imaging sensor
  using INTPIX4-SOI pixelated silicon devices}, Nuclear Instruments and Methods
  in Physics Research Section A: Accelerators, Spectrometers, Detectors and
  Associated Equipment 979 (2020) 164400.
\newblock \href {http://arxiv.org/abs/2006.05658} {\path{arXiv:2006.05658}},
  \href {https://doi.org/10.1016/j.nima.2020.164400}
  {\path{doi:10.1016/j.nima.2020.164400}}.

\bibitem{int4Manual}
Y.~Arai, {INTPIX4 User's Manual},
  https://soipix.jp/content/files/documents/INTPIX4doc\_v032.pdf Revision 0.33
  (2013).

\bibitem{int4Nishimura}
R.~Nishimura, Y.~Arai, T.~Miyoshi, K.~Hirano, S.~Kishimoto, R.~Hashimoto,
  {Development of an X-ray imaging system with SOI pixel detectors}, Nuclear
  Instruments and Methods in Physics Research Section A: Accelerators,
  Spectrometers, Detectors and Associated Equipment 831 (2016) 49--54.
\newblock \href {https://doi.org/10.1016/j.nima.2016.04.036}
  {\path{doi:10.1016/j.nima.2016.04.036}}.

\bibitem{int4Mitsui}
S.~Mitsui, Y.~Arai, T.~Miyoshi, A.~Takeda, {Development of integration-type
  silicon-on-insulator monolithic pixel detectors using a float zone silicon},
  Nuclear Instruments and Methods in Physics Research Section A: Accelerators,
  Spectrometers, Detectors and Associated Equipment 953 (2020) 163106.
\newblock \href {http://arxiv.org/abs/1804.03338} {\path{arXiv:1804.03338}},
  \href {https://doi.org/10.1016/j.nima.2019.163106}
  {\path{doi:10.1016/j.nima.2019.163106}}.

\bibitem{b10ZrS}
Z.~Wang, M.~Hoffbauer, C.~Morris, N.~Callahan, E.~Adamek, J.~Bacon, M.~Blatnik,
  A.~Brandt, L.~Broussard, S.~Clayton, C.~Cude-Woods, S.~Currie, E.~Dees,
  X.~Ding, J.~Gao, F.~Gray, K.~Hickerson, A.~Holley, T.~Ito, C.-Y. Liu,
  M.~Makela, J.~Ramsey, R.~Pattie, D.~Salvat, A.~Saunders, D.~Schmidt,
  R.~Schulze, S.~Seestrom, E.~Sharapov, A.~Sprow, Z.~Tang, W.~Wei, J.~Wexler,
  T.~Womack, A.~Young, B.~Zeck, {A multilayer surface detector for ultracold
  neutrons}, Nuclear Instruments and Methods in Physics Research Section A:
  Accelerators, Spectrometers, Detectors and Associated Equipment 798 (2015)
  30--35, b/Zn Detector.
\newblock \href {http://arxiv.org/abs/1503.03424} {\path{arXiv:1503.03424}},
  \href {https://doi.org/10.1016/j.nima.2015.07.010}
  {\path{doi:10.1016/j.nima.2015.07.010}}.

\bibitem{ucnTau21}
F.~M. Gonzalez, E.~M. Fries, C.~Cude-Woods, T.~Bailey, M.~Blatnik, L.~J.
  Broussard, N.~B. Callahan, J.~H. Choi, S.~M. Clayton, S.~A. Currie, M.~Dawid,
  E.~B. Dees, B.~W. Filippone, W.~Fox, P.~Geltenbort, E.~George, L.~Hayen,
  K.~P. Hickerson, M.~A. Hoffbauer, K.~Hoffman, A.~T. Holley, T.~M. Ito,
  A.~Komives, C.-Y. Liu, M.~Makela, C.~L. Morris, R.~Musedinovic,
  C.~O'Shaughnessy, R.~W. Pattie, J.~Ramsey, D.~J. Salvat, A.~Saunders, E.~I.
  Sharapov, S.~Slutsky, V.~Su, X.~Sun, C.~Swank, Z.~Tang, W.~Uhrich,
  J.~Vanderwerp, P.~Walstrom, Z.~Wang, W.~Wei, A.~R. Young, m.~x. w. o.~d.
  Collaboration, {Improved Neutron Lifetime Measurement with UCN$\tau$},
  Physical Review Letters 127~(16) (2021) 162501.
\newblock \href {http://arxiv.org/abs/2106.10375} {\path{arXiv:2106.10375}},
  \href {https://doi.org/10.1103/physrevlett.127.162501}
  {\path{doi:10.1103/physrevlett.127.162501}}.

\bibitem{ucnTau22}
C.~Cude-Woods, F.~M. Gonzalez, E.~M. Fries, T.~Bailey, M.~Blatnik, N.~B.
  Callahan, J.~H. Choi, S.~M. Clayton, S.~A. Currie, M.~Dawid, B.~W. Filippone,
  W.~Fox, P.~Geltenbort, E.~George, L.~Hayen, K.~P. Hickerson, M.~A. Hoffbauer,
  K.~Hoffman, A.~T. Holley, T.~M. Ito, A.~Komives, C.-Y. Liu, M.~Makela, C.~L.
  Morris, R.~Musedinovic, C.~O'Shaughnessy, R.~W. Pattie, J.~Ramsey, D.~J.
  Salvat, A.~Saunders, E.~I. Sharapov, S.~Slutsky, V.~Su, X.~Sun, C.~Swank,
  Z.~Tang, W.~Uhrich, J.~Vanderwerp, P.~Walstrom, Z.~Wang, W.~Wei, A.~R. Young,
  {Fill and dump measurement of the neutron lifetime using an asymmetric
  magneto-gravitational trap}, Physical Review C 106~(6) (2022) 065506.
\newblock \href {http://arxiv.org/abs/2205.02323} {\path{arXiv:2205.02323}},
  \href {https://doi.org/10.1103/physrevc.106.065506}
  {\path{doi:10.1103/physrevc.106.065506}}.

\end{thebibliography}

%============================
%============================
%============================

%============================
%============================
%============================

%============================
%============================
%============================

%============================
%============================
%============================

%============================
%============================
%============================

\end{document}